\documentclass[12pt]{article}
\usepackage{amsmath,amssymb,color}
\usepackage{cite,epsf} 
\usepackage{pstricks}
\usepackage{color}
\usepackage{hyperref}
\usepackage{tikz}     
\usepackage{graphicx}
\usetikzlibrary{
decorations.pathmorphing 
}

\usepackage{graphicx,array} 
\newcommand{\be}{\begin{equation}}
\newcommand{\ee}{\end{equation}}
\newcommand{\beq}{\begin{equation}}
\newcommand{\eeq}{\end{equation}}
\newcommand{\bea}{\begin{eqnarray}}
\newcommand{\eea}{\end{eqnarray}}

\newcommand{\ba}{\begin{eqnarray}}
\newcommand{\ea}{\end{eqnarray}}

\def\sin{\mbox{sin}}
\def\cos{\mbox{cos}}

\def\log{\mbox{log}}

\begin{document}

\begin{titlepage}
\vspace{10pt}
\hfill
{\large\bf HU-EP-20/01}
\vspace{20mm}
\begin{center}

{\Large\bf   On anomalous conformal Ward identities \\[2mm]
for Wilson loops on polygon-like contours \\[3mm]
with circular edges
}

\vspace{45pt}

{\large Harald Dorn 
{\footnote{dorn@physik.hu-berlin.de
 }}}
\\[15mm]
{\it\ Institut f\"ur Physik und IRIS Adlershof, 
Humboldt-Universit\"at zu Berlin,}\\
{\it Zum Gro{\ss}en Windkanal 6, D-12489 Berlin, Germany}\\[4mm]

\vspace{20pt}

\end{center}
\vspace{10pt}
\vspace{40pt}

\centerline{{\bf{Abstract}}}
\vspace*{5mm}
\noindent
We derive the anomalous conformal Ward identities for ${\cal N}=4$ SYM Wilson loops on polygon-like contours with edges formed by circular arcs. With a
suitable choice of parameterisation they are very similarly to those for local correlation functions. Their solutions have a  conformally
covariant factor depending on the distances of the corners times a conformally invariant remainder factor depending, besides on cross ratios of the corners, 
on the cusp angles and angles parameterising the torsion of the contours.

\vspace*{4mm}
\noindent

\vspace*{5mm}
\noindent
   
\end{titlepage}
\newpage


\section{Introduction}
Soon  after the renormalisation of pure Wilson loops or lines \cite{Polyakov:1980ca,Dotsenko:1979wb,Brandt:1981kf,Gervais:1979fv,Arefeva:1980zd},
the renormalisation and short distance properties of operator insertions into Wilson loops as well as of nonlocal gauge invariant gluonium and hadron
operators have been studied  \cite{Gervais:1979fv,  Dorn:1980hs,Craigie:1980qs,Dorn:1981wa,Dorn:1986dt,Polyakov:2000ti,Polyakov:2000jg }. In comparison to smooth contours, cusps or self-crossings of the contour require additional renormalisation. Using a one-dimensional auxiliary field living on the contour
\cite{Gervais:1979fv,Arefeva:1980zd,Dorn:1986dt}, one can formulate both the presence of cusps  and self-crossings as well of the insertion of additional operators in the language of correlation functions of local composites on the contour. 

Besides their importance for  the formal theory,  operator insertions into Wilson loops play  a crucial  role also in the theory of parton distributions one needs in phenomenology.\footnote{Instead giving a list of references we cite only two recent papers \cite{Wang:2019tgg,Radyushkin:2019mye} and refer to further citations therein.}

In particular since the invention of AdS-CFT duality \cite{Maldacena:1997re}, a huge amount of activity has been devoted to the ${\cal N}=4$ supersymmetric Yang Mills theory with its high degree of symmetry as superconformal invariance, integrability and access to the strong coupling regime. One aspect related to the present paper concerns the study of operator insertions in the Maldacena-Wilson loop \cite{Maldacena:1998im} along straight lines or circles 
\cite{Drukker:2006xg,Cooke:2017qgm,Giombi:2017cqn,Beccaria:2019dws,Liendo:2018ukf}. A complementary setting is connected with the duality
between scattering amplitudes and Wilson loops \cite{Alday:2007hr}. Here one needs Wilson loops for polygons with light-like edges. One has to study
not a given fixed contour with varying insertion points on it, but a contour determined by its special points, the corners of a light-like polygon.

Only recently a generalisation to another case with a Wilson loop contour fixed by the position of its cusp points plus a finite number of parameters 
has been studied, a triangle with circular edges \cite{Cavaglia:2018lxi,McGovern:2019sdd} in a plane. As a result they get  a dependence on the three
corner points as for a three point correlator of local composites with conformal dimensions equal to the cusp anomalous dimensions. Instead of the structure constant for the three local operators there appears a three cusp structure function, which is a new function of the three cusp angles.

The aim of our paper is to fully exploit conformal properties to constrain the structure of Wilson loops for arbitrary  polygon-like contours
with edges made of circular arcs, either in full 4D Euclidean spacetime or in two- or three-dimensional Euclidean subspaces of 4D Minkowski space.
We consider ${\cal N}=4$  SYM, but leave it open to have in mind either  ordinary Wilson  loops or their supersymmetric extensions  \cite{Maldacena:1998im}. The resulting structure will be the same, only the relevant cusp anomalous dimensions differ.
Our strategy will closely follow \cite{ Drummond:2007au} in deriving the appropriate anomalous conformal Ward identities. While for the light-like
 polygons studied in \cite{ Drummond:2007au} only cross ratios formed out of corner points are available as conformal invariant parameters, we will have to identify and handle a larger set of conformal invariants. 

The paper is organised as follows. In section 2 we derive the anomalous conformal Ward identities for our circular $N$-gons. For large enough $N$ it will turn out that the 
number of metrical parameters and the number of conformal parameters, needed in addition to those for the set of corners, are equal. Section 3
is devoted to this generic case. The exceptional cases  $N=2,3,4$ are treated in Section 4. We conclude with section 5 and have put various technical details in 
four appendices.
\section{Derivation of the anomalous conformal Ward identities}
The contours under consideration are polygon-like with edges made of circular arcs. Their geometry is fixed by the set of corner points (vertices)  $\{x_j\}$ and the set of centers for the circular arcs  $\{z_j\},~j=1\dots N$. We will call the transition from the set of corners to the complete circular $N$-gon as dressing. There is a constraint
\beq
(x_j-z_j)^2~=~(x_{j+1}-z_j)^2~,\label{rad-constr}
\eeq
and  the radii of the arcs
are given by 
\beq
R_j=\vert x_j-z_j\vert~.
\eeq 

We start with the anomalous Ward identities for dilatations
and special conformal transformations for Wilson loops in dimensionally regularised ${\cal N}=4$ SYM as derived in \cite{Drummond:2007au}
\bea
    {\cal D}~\log \langle{\mathbf W}_{\epsilon}\rangle&=&-\frac{2i\epsilon}{g^2\mu^{2\epsilon}}\int dx^{4-2\epsilon}~\frac{\langle{\cal L}(x){\mathbf W}_{\epsilon}\rangle}{\langle{\mathbf W}_{\epsilon}\rangle}~,\label{dil}\\
    {\cal K}^{\nu}~\log \langle{\mathbf W}_{\epsilon}\rangle&=&-\frac{4i\epsilon}{g^2\mu^{2\epsilon}}\int dx^{4-2\epsilon}~x^{\nu}~\frac{\langle{\cal L}(x){\mathbf W}_{\epsilon}\rangle}{\langle {\mathbf W}_{\epsilon}\rangle}~.\label{spec-conf}
    \eea

    For smooth contours the integrals on the r.h.s. of both equations are finite for $\epsilon\rightarrow 0$. Then the whole r.h.s. vanishes in the limit and restores conformal invariance. In the presence of cusps between spacelike edges the integrals have simple poles in $\epsilon$, and there remain anomalous terms on the r.h.s..

The further analysis in  \cite{Drummond:2007au} is tailored for polygons with lightlike edges
    where the presence of simple {\it and} double poles in $\log{\mathbf W}_{\epsilon}$ requires a more involved discussion.

    Instead we can use \footnote{$\Gamma_j=\Gamma(\alpha_j,g)$ denotes the cusp anomalous dimension for the cusp with opening angle $\alpha_j$ at $x_j$.}
\bea
\log \langle{\mathbf W}_{\epsilon}\rangle&=&\log Z~+~\log \langle{\mathbf W}\rangle ~,\label{Wren}\\
\log Z&=&-\frac{1}{\epsilon}~\int\frac{\gamma(g)}{g}dg~, ~\mbox{i.e.}~{\cal D}~\log Z=0~,~{\cal K}^{\nu}~\log Z=0~,\label{Z}\\
\gamma (g)&=&\sum_{j=1}^N\Gamma_j~,
\eea
and continue with a combination of the standard RG equation\footnote{ We have vanishing $\beta$-function in ${\cal N}=4$ SYM.}
    and dimensional analysis
\beq
{\cal D}~\log \langle{\mathbf W}\rangle~=~-\sum_{j=1}^N\Gamma_j~.
\eeq
Comparing this with \eqref{dil}, using \eqref{Wren},\eqref{Z}, and taking into account that the UV divergent contribution to the integral is located at the points $x=x_j$, one gets in the limit $\epsilon\rightarrow 0$
\beq
\sum_{j=1}^N\Gamma_j~\delta(x-x_j)~=~\frac{2i\epsilon}{g^2\mu^{2\epsilon}}~\frac{\langle{\cal L}(x){\mathbf W}_{\epsilon}\rangle}{\langle{\bf W}_{\epsilon}\rangle }~.\label{Gamma-delta}
\eeq
 For the renormalised Wilson loop this implies with \eqref{Z} after insertion of \eqref{Gamma-delta} into \eqref{spec-conf}
\beq
    {\cal K}^{\nu}~\log \langle{\mathbf W}\rangle~=~-2\sum_{j=1}^Nx_j^{\nu}~\Gamma_j~.\label{K-loop}
    \eeq

    The advantage of this equation for practical applications crucially depends on the choice of variables used to specify the contour of the Wilson loop. The first naive
    possibility would be
    \beq
    \langle{\mathbf W}\rangle~=~W(\{x\},\{z\})~.
    \eeq
    For the corner points $x_j$ we can use the standard infinitesimal transformation\\
    $({\cal K}^{\nu}x)^{\mu}=2x^{\nu}x^{\mu}-x^2\delta^{\mu\nu}$. However, in generic cases, under conformal transformations the center of the image of a circle is not the image of the original center. A detailed discussion of this issue one can find in  appendix A. From there, see  \eqref{trafo-z}, we get for a center of a circle
    \beq
    ({\cal K}^{\nu}z)^{\mu}~=~2z^{\nu}z^{\mu}-z^2\delta^{\mu\nu}-2R^2n^{\mu}n^{\nu}+R^2\delta^{\mu\nu}~,
    \eeq
    where $R$ and $n$ are radius and unit normal direction to the plane in which the circle is located.\footnote{This is for 3D. In 4D there appear two normals.}

    Then \eqref{K-loop} appears as
    \bea
    \sum_{j=1}^N\Big (2x_j^{\nu}x_j\partial_{ x_j}-x_j^2\partial_{ x_j}^{\nu} +2z_j^{\nu}z_j\partial_{ z_j}-(z_j^2-R_j^2)\partial_{ z_j}^{\nu}-2R_j^2n_j^{\nu}n_j\partial_{z_j} \Big)~\log W(\{x\},\{z\})\nonumber
    \\
    =~-2 \sum_{j=1}^Nx_j^{\nu}~\Gamma_j~.~~~~~~~~~~~~~~~~~~~~~~~~~~~~~~~~~~~~~~~~~
    \eea
    It is hard to disentangle some explicit constraints on the
    function $W(\{x\},\{z\})$  out of this version.

    As further input, as in the case of local correlation functions, one should use
    unbroken Poincar$\acute{\mbox{e}}$ invariance. The metrical properties of the set of corners depend only on the distances
    \beq
    D_{ij}~=~\vert x_i-x_j\vert ~,
    \eeq
    and one can apply the well-known
    \beq
        {\cal K}^{\nu}~D_{ij}~=~(x_i^{\nu}+x_j^{\nu})~D_{ij}~.\label{K-D}
        \eeq
Then a complete set of variables for our Wilson loops, which in 3D  immediately intrudes oneselves, is $\{D_{ij},R_j,\alpha_j\}$.  The cusp angles $\alpha_j$ are invariant. The transformation law of the $D_{ij}$   contains only the distances themselves and the corners $x_j$, which anyway are needed to fit their appearance on the r.h.s. of \eqref{K-loop}. But the transformation of the radii, see appendix A \eqref{trafo-R}, contains the centers of the circles which are
fixed by a nontrivial function of $D_{ij},R_j,\alpha_j$. 

Let us look in the next section for a more convenient choice of variables.
\section{Anomalous conformal Ward identities in the \\generic case $N\geq D+1$}
In appendix B we present a detailed counting of variables fixing with respect to their  metrical and conformal properties both  the set of
corners $\{x_j\}$ as well as our complete polygons $\{x_j,z_j\}$. As found in \eqref{delM}, the number of variables needed for dressing the set of corners to the full circular $N$-gon is both in the metric as well as the conformal case equal to $N(D-1)$. Therefore, we should parameterize our Poincar$\acute{\mbox{e}}$ invariant 
Wilson loop from the start  by the distances between the corner points and the $N(D-1)$ conformal dressing invariants. Clearly, $N$ of these
invariants are the cusp angles $\alpha_j$. The other $N(D-2)$ conformal invariants are torsion angles $\beta_{a,j},~a=1,\dots, D-2,~j=1,\dots, N$. They will be specified below.

For the moment we continue with
\beq
\langle {\mathbf W}\rangle~=~W(\{x_j\},\{z_j\})~=~{\cal W}(\{D_{ij}\},\{\alpha_j\},\{\beta_{a,j}\})~
\eeq
and get from \eqref{K-loop} and \eqref{K-D}
\beq
\sum_{i<j} \big(x_i^{\nu}+x_j^{\nu}\big )D_{ij}\frac{\partial}{\partial D_{ij}}~\log{\cal W}(\{D_{ij}\},\{\alpha_j\},\{\beta_{a,j}\})~=~ -2 \sum_{j=1}^Nx_j^{\nu}~\Gamma_j~.\label{master}
\eeq
The general solution of this equation is \footnote{We suppress here and below a factor $\mu^{\sum\Gamma_j}$ necessary to fit the correct engineering dimension (with $\mu$ as RG scale).}
\beq
\langle{\mathbf W}\rangle ~=~{\cal W}~=~\Omega \big(\{c\},\{\alpha\},\{\beta\}\big)~\prod _{i<j}D_{ij}^{m_{ij}}, \label{calW}
  \eeq
  where $\Omega$ is a function of the cusp and torsion angles as well as of the independent cross ratios $\{c\}$ formed out of the set $\{D_{ij}\}$. The $m_{ij}$ depend only on the cusp angles $\{\alpha\}$
\footnote{According to common belief the cusp anomalous dimension depends on the cusp angle only. In weak coupling perturbation theory
this is well established. Using AdS/CFT it has been proven at strong coupling for the general planar case \cite{Dorn:2015bfa}. To my knowledge there is still lacking an explicit proof in the presence of torsion.}
 and
  have to obey $m_{ij}=m_{ji}$ and
  \beq
  \sum _{j\neq i} m_{ij}~=~-2\Gamma_i~.\label{m-Gamma}
  \eeq

  We found a structure very similar to that for correlation functions of local operators with conformal dimension $\Gamma_j$ at points $x_j$. There the remainder factor multiplying
  the powers of the $D_{ij}$ depends only on the cross ratios. Here this factor depends in addition on the cusp and torsion angles, but on nothing else.\\

  We still have to explain the torsion angles. In the planar case $D=2$ the metric structure is fixed by the corner distances $D_{ij}$ and the cusp angles. Going to higher dimensions, the circular polygon can wind out of a given plane. Concerning the metrical issues one could this describe by angles relative to planes. But under conformal
  transformations  planes are not mapped to planes in general. To choose from the beginning additional variables which are conformally invariant, one has to rely on angles to circles and spheres.

\begin{figure}[h!]
\begin{center}
 \includegraphics[width=10cm]{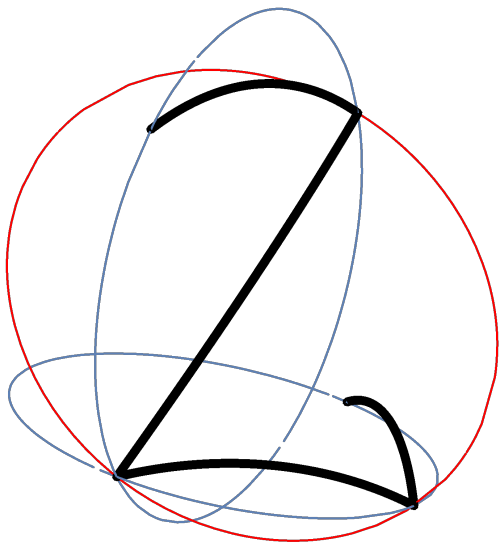}
\end{center}
\caption {\it In black is shown a generic piece of our contours (between corners $x_{j-2},x_{j-1},x_j,x_{j+1},x_{j+2}$) .  The circumcircle $\{x_{j-1},x_j,x_{j+1}\}$ appears in red and the circumcircles $ \{x_{j-2},x_{j-1},x_j\}$ and $\{x_j,x_{j+1},x_{j+2}\}$ in blue.}
\label{fig:generic-piece}.
\end{figure}

\begin{figure}[h!]
\begin{center}
 \includegraphics[width=8cm]{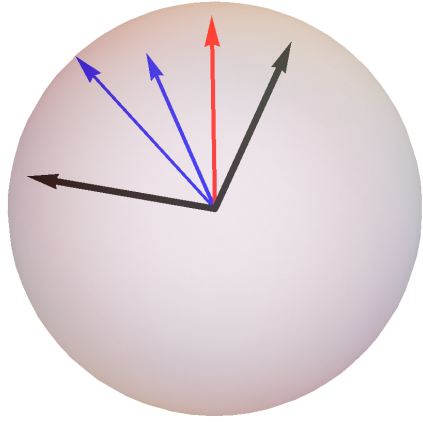}
\end{center}
\caption {\it The unit sphere at a corner $x_j$ with five directions corresponding to the three circumcircles and the two polygon edges as in the previous figure.
Note that the numerics of both figures is not tuned.} 
\label{fig:directions}
\end{figure}

  Let us discuss the case $D=3$ in detail and make some comments on $D=4$ afterwards.
  In fig.\ref{fig:generic-piece} we present a generic piece of a circular polygon
  between five consecutive corners and circumcircles for three triangles with
  corners, which are consecutive on the polygon. Since circles are mapped to circles, all angles at the central corner of fig.\ref{fig:generic-piece} are conformally invariant.
What are the correlations between these angles and which ones are sufficient to
describe the dressing of the corner set by the circular edges? In fig.\ref{fig:directions} we show the unit tangent vectors of the three circumcircles and the two
circular edges meeting at $x_j$. The circumcircles are fixed by the corners. To fix in addition
the circular edge $\{x_j,x_{j+1}\}$, it is sufficient to know the direction of its tangent relative that of the circumcircles. From the metrical data of the
set of corners one can reconstruct the angle to the straight line between $x_j$ and $x_{j+1}$ which fixes the radius of the circular edge. Together with the known direction this fixes then also the position of  its center $z_j$.

Let us introduce a short hand notation for the circumcircles $cc_j=\{x_{j-1},x_j,x_{j+1}\},$ $cc_j^+=\{x_{j},x_{j+1},x_{j+2}\},$ $cc_j^-=\{x_{j-2},x_{j-1},x_{j}\}$. Now we start at $x_j$ and fix the edge $\{x_{j-1},x_j\}$ by the angles $\measuredangle (\{x_{j-1},x_j\},cc_j)$ and  $\measuredangle (\{x_{j-1},x_j\},cc^-_j)$. 

Then we fix the edge $\{x_j,x_{j+1}\}$ by the angles 
\bea
\alpha_j&=&\measuredangle (\{x_{j},x_{j-1}\},\{x_{j},x_{j+1}\})~,\\
\beta_j&=&\measuredangle (\{x_{j},x_{j+1}\},cc_j)~.\label{betaj}
\eea
In this manner we continue along the circular polygon, fixing all edges by the cusp angles $\alpha_j$ and the torsion angles $\beta_j$, defined by \eqref{betaj}
at each corner $x_j$ as the angle to the corresponding red circumcircle $cc_j$. Reaching corner $x_{j-1}$ we have to remember that the edge $\{x_{j-1},x_j\}$
has been fixed already at the beginning of our tour along the polygon. But from fig.\ref{fig:generic-piece} it is obvious that 
\bea
\measuredangle (\{x_{j},x_{j-1}\},cc_j)&=&\measuredangle (\{x_{j-1},x_j\},cc^+_{j-1})~,\label{anglesxj}\\
\measuredangle (\{x_{j},x_{j-1}\},cc^-_j)&=&\measuredangle (\{x_{j-1},x_j\},cc_{j-1})~=~\beta_{j-1}~.
\eea 
Since at the corner $x_{j-1}$ the r.h.s. of \eqref{anglesxj} can be expressed in terms of $\alpha_{j-1}$ and  $\beta_{j-1}$, we altogether have
shown that our circular polygons,  up to  Poincar$\acute{\mbox{e}}$ transformations, are fixed by the set of variables $\{D_{ij},\alpha_j,\beta_j\}$, confirming \eqref{calW}.

Concerning the  conformal analysis of contours, we have found in the mathematical literature only papers on smooth contours, see \cite{Cairns} and references therein. In analogy to the description of metrical invariants  by the Frenet formulas, they introduce conformal length, conformal curvature and
conformal torsion, see appendix C. In their language an  edge of our polygons is a conformal vertex, the conformal length stays constant. It would be interesting to fit our conformal description of piecewise smooth contours to that of smooth contours by allowing distributional conformal curvature etc.

We close this section with a comment on $D=4$. While, as just discussed, in $D=3$ the direction of one of the black arrows in fig.\ref{fig:directions}  is fixed by the angle to the other
black arrow $( \alpha_j)$ and the angle to the red arrow $(\beta_j=\beta_{1,j})$, in $D=4$ one needs in addition the angle to one of the blue arrows $(\beta_{2,j})$. 
\section{The exceptional cases $N<D+1$}
According to the tables in appendix B, the circular two-gon has three metrical invariants in $D=2$, and four in higher dimensions. We take $D_{12},R_1,R_2$ and the cusp
angle $\alpha$, understanding that in a plane $\alpha $ is fixed by the other three
variables. The set of the two corners defines no plane or circumcircle as used
for higher N-gons to describe torsion. In all dimensions our two-gons have one conformal invariant, it is of course the cusp angle $\alpha=\alpha_1=\alpha_2$.

Due to  Poincar$\acute{\mbox{e}}$ invariance we can start from
\beq
\langle {\mathbf W_2}\rangle~=~W_2(x_1,x_2,z_1,z_2)~=~{\cal W}_2(D_{12},R_1,R_2,
\alpha)~
\eeq
and get with \eqref{K-D}  and \eqref{trafo-R} as anomalous conformal Ward identity
(note $\Gamma_1=\Gamma_2=\Gamma$)
\beq
\Big (\sum_{j=1,2}2z_j^{\nu}R_j\frac{\partial}{\partial R_j}+(x_1^{\nu}+x_2^{\nu})D_{12}\frac{\partial}{\partial D_{12}}\Big)\log~ {\cal W}_2~=~-2(x_1^{\nu}+x_2^{\nu})~ \Gamma~.\label{ward2D}
\eeq
The solution is $\log {\cal W}_2=-2\Gamma \log D_{12}+\log~ \omega_2$ where $\log~\omega_2$
now obeys the homogeneous variant of \eqref{ward2D}. Using translation invariance
to put $z_2=0$  one gets
\beq
\Big(2z_1^{\nu}R_1\frac{\partial}{\partial R_1}+(x_1^{\nu}+x_2^{\nu})D_{12}\frac{\partial}{\partial D_{12}}\Big )\log~\omega_2(D_{12},R_1,R_2,\alpha)~=~0~.\label{conf-omega2}
\eeq
In a generic case in $D>2$, rotation invariance can be used to get e.g. $z_1^1=0$ and $x_1^1+x_2^1\neq 0$ or vice versa. This implies that $\omega_2$ does not depend
on $D_{12}$ and $R_1$, and then for symmetry reasons also not on $R_2$.

In the case $D=2$ the points $z_1, (x_1+x_2)/2,z_2$ are located on a straight line, thus obstructing the argument just given. But then, since $\alpha$ is fixed by $D_{12},R_1,R_2$, the function $\omega_2$ can be treated from the beginning as $\omega_2(D_{12},R_1,\alpha)$. Dimensional analysis then yields $D_{12} \partial_{D_{12}}=-R_1\partial_{R_1}$, and in combination with \eqref{conf-omega2} we get for $\omega_2$ again independence of $D_{12}$ and of the radii.

In conclusion we found for the circular two-gon in arbitrary dimensions
\beq
\langle {\mathbf W}_2\rangle~=~{\cal W}_2~=~D_{12}^{-2\Gamma}~\omega_2(\alpha)~.
\eeq
\\

The circular triangle in $D=2$ still belongs to the generic case of the previous section. There are three conformal invariants, the cusp angles., hence we get
\beq
{\cal W}_3~=~D_{12}^{\Gamma_3-\Gamma_1-\Gamma_2}D_{23}^{\Gamma_1-\Gamma_2-\Gamma_3}D_{13}^{\Gamma_2-\Gamma_1-\Gamma_3}~\Omega_3(\alpha_1,\alpha_2,\alpha_3)~,~~\mbox{for}~D=2~.
\eeq

For a detailed discussion of the conformal triangle geometry in $D=3$ we have added appendix D. From there, as well from the tables in appendix B,
we know 5 invariants, i.e. 3 cusp and 3 torsion angles with one closing constraint \eqref{closure}. For fixing the Poincar$\acute{\mbox{e}}$ invariant
structure, we need 6 variables in addition to the distances of the corners. Then, similar to the two-gon discussion above, we have to study a function
of the distances and e.g. 
$R_1,\alpha_1,\alpha_2,\alpha_3,\beta_1,\beta_2$, which solves the homogeneous (unbroken) version of the Ward identity
\beq
\Big(2z_1^{\nu}R_1\partial _{R_1}+\sum_{i<j}(x_i^{\nu}+x_j^{\nu})D_{ij}\partial_ {D_{ij}}\Big )\log~\omega_3(\{D_{ij}\},R_1,\alpha_1,\alpha_2,\alpha_3,\beta_1,\beta_2)~=~0~.\label{conf-omega3}
\eeq
After a translation by $-z_1$ the derivative w.r.t. $R_1$ no longer contributes, and we have a homogeneous system of  three linear equations for
the derivatives w.r.t. the distances. Its coefficient determinant is equal to the scalar triple product \\
$(x_1+x_2-2z_1)\cdot\big((x_2+x_3-2z_1)\times (x_1+x_3-2z_1)\big)$ and generically different from zero. Hence the derivatives w.r.t. the distances
are all zero. Using this in the unshifted version of \eqref{conf-omega3} one concludes, that also the derivative of $\omega_3$  w.r.t. $R_1$ is zero.
As a consequence this yields the structure
\beq
{\cal W}_3~=~D_{12}^{\Gamma_3-\Gamma_1-\Gamma_2}D_{23}^{\Gamma_1-\Gamma_2-\Gamma_3}D_{13}^{\Gamma_2-\Gamma_1-\Gamma_3}~\omega_3(\alpha_1,\alpha_2,\alpha_3,\beta_1,\beta_2)~,~~\mbox{for}~D=3~.
\eeq

In a similar way, for a  circular triangle in $D=4$ we have to handle the gap between 6 conformal invariants and 8 metrical variables for dressing the
set of corners. Then one can argue that the derivatives of  $\hat \omega_3(\{D_{ij}\},R_1,R_2,\alpha_1,\alpha_2,\alpha_3,\beta_1,\beta_2,\beta_3)$ 
w.r.t. the $D_{ij},R_1$ and $R_2$ are zero if the vectors $z_2-z_1,~x_1+x_2-2z_1,~x_2+x_3-2z_1,~x_1+x_3-2z_1$ are linearly independent, i.e. in the generic 4$D$ case.
Hence we get
\beq
{\cal W}_3~=~D_{12}^{\Gamma_3-\Gamma_1-\Gamma_2}D_{23}^{\Gamma_1-\Gamma_2-\Gamma_3}D_{13}^{\Gamma_2-\Gamma_1-\Gamma_3}~\hat \omega_3(\alpha_1,\alpha_2,\alpha_3,\beta_1,\beta_2,\beta_3)~,~~\mbox{for}~D=4~.
\eeq
\\

In dimensions up to four, the circular tetragon in $D=4$ is the last exceptional case. According to the last table in appendix B one needs 12 metrical invariants
for dressing, but has only 11 conformal invariants. Similar to the triangle case in $D=3$ there is obviously a constraint among the angles
\beq
F(\alpha_1,\dots, \alpha_4,\beta_{1,1},\dots,\beta_{1,4},\beta_{2,1},\dots ,\beta_{2,4})~=~0~.
\eeq
One gets for $D=4$
\beq
{\cal W}_4~=~\prod _{1\leq i<j\leq 4}D_{ij}^{m_{ij}}~ \omega_4\big (\{c\},\{\alpha_{j=1,\dots,4}\},\{\beta_{1,j=1,\dots,4}\},\{\beta_{2,j=1,\dots,3}\}\big )~,
\eeq
with $m_{ij}$ obeying \eqref{m-Gamma} and $\{c\}=\big\{\frac{D_{12}^2D_{34}^2}{D_{13}^2D_{24}^2},\frac{D_{14}^2D_{23}^2}{D_{13}^2D_{24}^2}\big\}$.
\section{Conclusions}
 Our main result are the anomalous conformal Ward identities for Wilson loops on polygons with circular edges in the form of \eqref{master}. Due to skilfully chosen parameterisation they
look exactly like the corresponding identities for usual local correlation functions of composite operators with given conformal dimensions sitting
just on the points $x_j$, where the corners of our Wilson loop contours are located. Then the  solutions to these identities  have the same structure.
They are given in both cases by a conformally covariant factor composed out of powers of distances $\vert x_i-x_j\vert$ times a conformally invariant remainder factor, depending
in the case of local correlation functions on the cross ratios formed out of the $x_j$, but depending in the case of the Wilson loops on the same  cross ratios 
{\it and}
cusp and torsion angles.

While  the general strategy for the derivation of the  identities followed standard reasoning and \cite{Drummond:2007au}, on the technical level we had to
 make some effort to find the most convenient set of variables to parameterise our polygons. Besides the obvious distances between the corner points and the cusp angles we
found a way to describe the contours torsion freedom. We did it via angles between the edges and circumcircles fixed by the corresponding corner points and their neighbours. We also
discussed subtleties concerning the number of all these variables in dependence on the dimension $D$ and the number of corners $N$.

Our detailed discussion concerned only Wilson loops where nonzero conformal dimensions are generated by the UV renormalisation enforced by the presence
of cusps. However, it is straightforward to apply our formulas also to the case where at the locations of the cusps additional local operators are
inserted. Then the $\Gamma_j$ in \eqref{master} no longer stand for the cusp anomalous dimension $\Gamma (\alpha_j)$, but for
the complete (engineering + quantum correction) conformal dimension $\Gamma_j(\alpha_j)$ of the operator number $j$, dressed by the presence
of the Wilson loop.

For further work it would be interesting to identify dependence of the remainder factor on torsion angles by explicit calculations. Another interesting issue
is the generalisation of our discussion to Maldacena-Wilson loops which, in addition, at the corners have a discontinuity in their coupling to the scalars. Then one has to parameterise some kind of  torsion in $S^5$ and to fully exploit superconformal invariance analysis.
\\[10mm]
{\bf Acknowledgement:}\\
I thank George Jorjadze for useful discussions and the Quantum Field and String Theory group at Humboldt University for kind hospitality.
\section*{Appendix A: Conformal maps of circular arcs} 
Here we discuss the mapping of a circular arc with radius $R$ between two points $x_1$ and $x_2$  under a special conformal transformation
\beq
x\mapsto y~=~\frac{x + c~ x^2}{1+2cx+c^2x^2}~.\label{map}
\eeq
To fix the location of the arc, one still has to specify the plane in which the circle is embedded. The plane is fixed by $x_1,x_2$ and the center of the circle
$z$. This center is constrained by
\beq
(x_1-z)^2~=~R^2~=~(x_2-z)^2~,\label{constraint1}
\eeq
and there is of course another constraint on $R$
\beq
(2R)^2~>~(x_1-x_2)^2~.\label{constraint2}
\eeq
The images of the endpoints of the circular arc are given by applying the map \eqref{map} to $x_1$ and $x_2$, respectively. However, the center of the full circle,  of which the image of our arc is a certain part,  {\it is not} given by the image of  $z$ under \eqref{map}.  Instead we have to construct the image circle
out of the intersection of the image of the sphere
\beq
(x-z)^2~=~R^2\label{sphere}
\eeq
and the image of the plane
\beq
xn~=zn~,~~~~\mbox{with}~~~n=(x_1-z)\times(x_2-z)~.\label{plane}
\eeq
The image of the sphere \eqref{sphere} is another sphere given by
\bea
(y-y_A)^2&=&R_A^2~,\label{sphere-image}\\
y_A&=&\frac{z+c(z^2-R^2)}{1+2cz+c^2(z^2-R^2)}~,\nonumber\\
R_A^2&=&\frac{R^2}{(1+2cz+c^2(z^2-R^2))^2}~.\nonumber
\eea
The image of the plane \eqref{plane}
is a sphere
\bea
(y-y_B)^2&=&R_B^2~,\label{plane-image}\\
y_B&=&\frac{1}{2}~\frac{n+2c(nz)}{nc+c^2(nz)}~,\nonumber\\
R_B^2&=&\frac{n^2}{4(nc+c^2(nz))^2}~.\nonumber
\eea
Then the radius $\hat R$ of the image circle is given as a solution of
\beq
\sqrt{R_A^2-\hat R^2}~+~\sqrt{R_B^2-\hat R^2}~=~\vert y_A-y_B\vert
\eeq
and its center $\hat z$ by
\beq
\hat z~=~y_A~+~\frac{\sqrt{R_A^2-\hat R^2}}{\vert y_A-y_B\vert}~(y_B-y_A)~.
\eeq
Inserting the information contained in \eqref{sphere-image} and \eqref{plane-image} we get
\beq
\hat z=\frac{2R^2(nc+c^2(nz))(n+2c(nz))+n^2(1+2cz+c^2(z^2-R^2))(z+c(z^2-R^2))}{4R^2(nc+c^2(nz))^2+n^2(1+2cz+c^2(z^2-R^2))^2}\label{y0}
\eeq
and
\beq
\hat R^2=\frac{R^2n^2}{4R^2(nc+c^2(nz))^2+n^2(1+2cz+c^2(z^2-R^2))^2}~.\label{Rtilde}
\eeq
Note that in the special case $nc=nz=0$ the plane \eqref{plane} is mapped to itself, resulting in  $\hat z=y_A,~\hat R=R_A$. \footnote{There is another special case if the image of the sphere \eqref{sphere} is a plane. Then one has  \\$~~~~~~~\hat z=y_B,~\hat R=R_B$.}

For the application to conformal Ward identities we need \eqref{y0} and \eqref{Rtilde}  for infinitesimal $c$ only
\bea
\hat z&=&z+cz^2-2(cz) z+\frac{2R^2(cn)n}{n^2}-R^2 c~+~{\cal O}(c^2)~,\label{trafo-z}\\
\hat  R&=&R(1-2cz)~+~{\cal O}(c^2)~.\label{trafo-R}
\eea
\section*{Appendix B: Counting of metrical and conformal\\ parameters}
Let us start with the well-known counting of metrical invariants for the set \\
$\{x_j\},~j=1,\dots,N$ in $D$ dimensions. The number of generators of the Poincar$\acute{\mbox{e}}$  group is
$\frac{D(D+1)}{2}$, hence one gets for large enough $N$ the number of metrical invariants as $ND-\frac{D(D+1)}{2}$.
However, there is a little subtlety. A generic set of $N$ points spans a $(N-1)$-dimensional space. Then the last expression yields $\frac{N(N-1)}{2}$. Putting
$N$ points in a space of larger dimension, the number of metrical invariants remains unchanged. Therefore, one gets altogether
\beq
M_{\mbox{\scriptsize met}}^{\{x\}}~=~\frac{N(N-1)}{2}~\Theta(D+1-N)~+~\Big(ND-\frac{D(D+1)}{2}\Big)~\Theta(N-D-1)~.\label{Mxm}
\eeq
To avoid double counting in the case where the arguments of the UnitStep functions are zero, $N$ should be always understood as $\lim_{\delta\rightarrow 0}(N+\delta)$.

For our polygon with circular edges the corresponding number is 
\beq
M_{\mbox{\scriptsize met}}^{\{x,z\}}~=~2N(N-1)~\Theta(D+1-2N)~+~\Big(2ND-N-\frac{D(D+1)}{2}\Big )~\Theta(2N-D-1)~.\label{Mxzm}
\eeq
The term $-N$ in the big bracket is due to the $N$  radius constraints \eqref{rad-constr}. 

The analogous numbers for counting the conformal invariants are
\beq
M_{\mbox{\scriptsize conf}}^{\{x\}}=\frac{N(N-3)}{2}\Theta(D+1-N)+\Big(ND-\frac{(D+1)(D+2)}{2}\Big )\Theta(N-D-1)~,\label{Mxc}
\eeq
\bea
M_{\mbox{\scriptsize conf}}^{\{x,z\}}&=&2N(N-2)\Theta(D+1-2N)\nonumber\\
&&+~\Big(2ND-N-\frac{(D+1)(D+2)}{2}\Big)\Theta(2N-D-1).\label{Mxzc}
\eea
Note also, that the  prefactors of the two UnitSteps agree for $N=D$ and $N=D+1$ in \eqref{Mxm}, for $2N=D,D+1$ in \eqref{Mxzm} as well as for $N=D+1$ and $N=D+2$ in \eqref{Mxc} and $2N=D+1,D+2$ in \eqref{Mxzc}.

For large enough $N$, both in the metrical as well as in the conformal case,
the difference in the number of invariants between the circular polygon and
the set of corners $\{x_j\}$ is equal to
\beq
\Delta M~=~N(D-1)~. \label{delM}
\eeq
For small $N$ there are deviations from this rule due to the presence of
various UnitStep terms. For some illustration see the following tables. From these tables we see, that deviations from the rule \eqref{delM}
hold at $N< D+1$ in the conformal case and at $N< D$ in the metrical case.
\\[10mm]
\noindent
$D~=~2$\\[2mm]
\begin{tabular}[h]{|c|c c c c c c}
    \hline
  $N$&2&3&4&5&6&\dots\\
  \hline
  $M_{\mbox{\scriptsize met}}^{\{x\}}$&1&3&5&7&9&\dots\\
  $M_{\mbox{\scriptsize met}}^{\{x,z\}}$&3&6&9&12&15&\dots\\
  $M_{\mbox{\scriptsize conf}}^{\{x\}}$&0&0&2&4&6&\dots\\
   $M_{\mbox{\scriptsize conf}}^{\{x,z\}}$&1&3&6&9&12&\dots\\
  \hline
 \end{tabular}\\[10mm]
$~~~~~~~~~~~~~~~~~~~~D~=~3$\\[2mm]
$~~~~~~~~~~~~~~~~~~~~$\begin{tabular}[h]{|c|c c c c c c}
  \hline
  $N$&2&3&4&5&6&\dots\\
  \hline
  $M_{\mbox{\scriptsize met}}^{\{x\}}$&1&3&6&9&12&\dots\\
  $M_{\mbox{\scriptsize met}}^{\{x,z\}}$&4&9&14&19&24&\dots\\
  $M_{\mbox{\scriptsize conf}}^{\{x\}}$&0&0&2&5&8&\dots\\
   $M_{\mbox{\scriptsize conf}}^{\{x,z\}}$&1&5&10&15&20&\dots\\
  \hline
\end{tabular}\\[10mm]
$~~~~~~~~~~~~~~~~~~~~~~~~~~~~~~~~~~~~~~~D~=~4$\\[2mm]
$~~~~~~~~~~~~~~~~~~~~~~~~~~~~~~~~~~~~~~~$\begin{tabular}[h]{|c|c c c c c c}
  \hline
  $N$&2&3&4&5&6&\dots\\
  \hline
  $M_{\mbox{\scriptsize met}}^{\{x\}}$&1&3&6&10&14&\dots\\
  $M_{\mbox{\scriptsize met}}^{\{x,z\}}$&4&11&18&25&32&\dots\\
  $M_{\mbox{\scriptsize conf}}^{\{x\}}$&0&0&2&5&9&\dots\\
   $M_{\mbox{\scriptsize conf}}^{\{x,z\}}$&1&6&13&20&27&\dots\\
  \hline
\end{tabular}\\[10mm]
Finally two remarks are in order. For $N=2$ formula \eqref{Mxc} yields a negative value, of course this has to be replaced by zero. Furthermore formula \eqref{Mxzc}
gives zero. Obviously this is wrong, since the cusp angle of a circular two-gon
is conformally invariant. The failure of our formula in this special case is due to the fact that there exist nontrivial conformal transformations which map
the circular two-gon to itself.
\section*{Appendix C: Conformal invariants of smooth\\ contours}
In the mathematical literature there are various papers on the conformal invariants of smooth contours, see \cite{Cairns} and references therein. In analogy to the metric invariants length $s$, curvature $\kappa$ and torsion $\tau$ they find \footnote{We present it for contours in $\mathbb R^3$. In higher dimensions there are more torsion invariants.}
conformal length $\omega$
\beq
d\omega~=~\sqrt{\nu}ds,~~~~~~\nu~=~\sqrt{(\kappa ')^2+\kappa ^2\tau ^2}~,
\eeq
conformal curvature $Q$
\beq
Q~=~\frac{4(\nu ''-\kappa ^2\nu)\nu -5(\nu ')^2}{8\nu ^3}~,
\eeq
and conformal torsion $T$
\beq
T~=~\frac{2(\kappa ')^2\tau+\kappa^2\tau ^3+\kappa\kappa'\tau '-\kappa\kappa''\tau}{\nu ^{\frac{5}{2}}}~.
\eeq
In this framework points where $\nu=0$ are called vertices, hence 
our edges are conformal vertices. Conformal curvature and torsion is localised at the
corners of our polygons, where smoothness is violated.
\section*{Appendix D: Conformal geometry of triangles with circular edges}
\begin{figure}[h!]
\begin{center}
 \includegraphics[width=8cm]{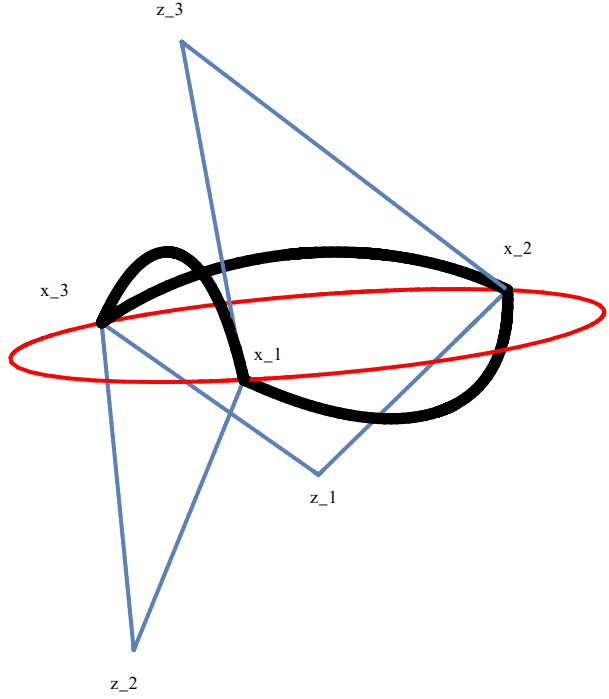}
\end{center}
\caption {\it A triangle with corners $x_1,x_2,x_3$ and edges out of circular arcs is shown in black. The corresponding circum circle is depicted in red. The blue lines are the radii connecting the corners with the centers related to the  circular arcs.  }
\label{fig:triangle}
\end{figure}

Here we want to discuss the conformal geometry of a triangle with circular edges in 3 dimensions. \footnote{In contrast to generic $N$-gons, following tradition in trigonometry, in this appendix we number the edges according to their opposite corner.\label{footnote-10}} As stated in the main text and in the appendix B, it has 5 conformal
invariants. Obviously, there has to be a constraint among the three cusp angles and the three torsion angles. The reason for this constraint can be seen
by  comparing fig.\ref{fig:generic-piece} and fig.\ref{fig:triangle}. In the case of a triangle there is only one circumcircle at our disposal.

 The angles $\alpha_j$ are the angles between the black
 circular arcs at the corners $x_j$. Following the convention stated in the footnote
 \ref{footnote-10}, we denote by $\beta_j$ the angle between the circular arc opposite to
the  corner $x_j$ and the red circum circle.

\begin{figure}[h!]
\begin{center}
 \includegraphics[width=10cm]{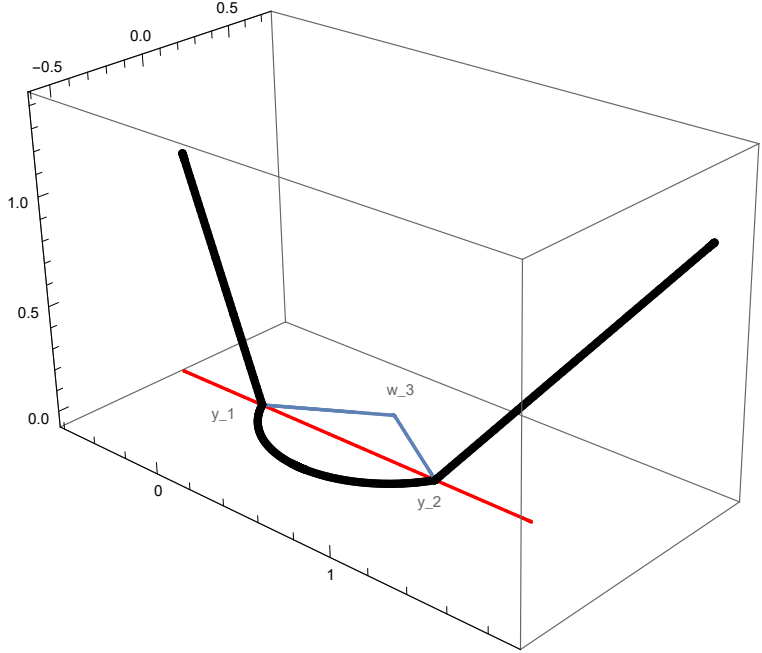}
\end{center}
\caption {\it The  image of a triangle in the conformal frame is shown in black.  $y_1,y_2$ are the images of the corners $x_1,x_2$. The image of $x_3$ is at  infinity.  The images of the edges number 1 and 2 are straight lines.
$w_3=\hat z_3$ denotes the center of the circle related to the edge number 3. The red line is the image of the circumcircle. The $\beta_j$'s are the angles between the red line and the black contour: $\beta_1$ and $\beta_3$ at $y_2$ and $\beta_3$ and $\beta_2$ at $y_1$. $\alpha_1$ and $\alpha_2$ are the angles between the pieces of the black contour at $y_1$ and $y_2$, respectively. $\vartheta_j,\varphi_j$ specify the direction of the straight line starting at $y_j$, \small{(j=1,2)}. }
\label{fig:conformal-frame}
\end{figure}

Let us still discuss an alternative  identification of  the independent conformal invariants.
We map a generic circular triangle to one in a conformal frame.  Such a frame can be defined by the conditions
\beq
y_1~=~0~,~~~y_2~=~(1,0,0)~,~~~y_3~=~\infty ~
\eeq
for the corners of the image  and the requirement that the center $\hat z_3$ of the circular arc between $y_1$ and $y_2$ is located in the (1,2)-plane, see fig. \ref{fig:conformal-frame}.

Under such a map the circular arcs connecting  $x_1$ and $x_2$ with $x_3$ are mapped to the straight lines passing $y_1$ and $y_2$, respectively.
Let the directions of these two straight lines be parameterised by $e_j=(\sin \vartheta_j\cos\varphi_j,\sin\vartheta_j\sin\varphi_j,\cos\vartheta_j)~,~~j=1,2$.
Then the image of a generic circular triangle in the conformal frame is fixed by the five parameters 
$$ \vartheta_1,~\vartheta_2,~\varphi_1,~\varphi_2~~\mbox{and}~\hat R_3~.$$
Instead of $\hat R_3$ we can use the angle $\beta_3$ between the circular arc and the straight line connecting $y_1$ and $y_2$. Their relation is
\beq
\hat R_3~=~\frac{1}{2~\sin\beta_3}~.
\eeq
The three angles $\alpha_j$ can be expressed via
\bea
\cos\alpha_1&=&\sin\vartheta_1~\cos(\beta_3+\varphi_1)~,\nonumber\\
\cos\alpha_2&=&-\sin\vartheta_2~\cos(\beta_3-\varphi_2)~,\nonumber\\
\cos\alpha_3&=&\cos\vartheta_1~\cos\vartheta_2~+~\sin\vartheta_1~\sin\vartheta_2~\cos(\varphi_1-\varphi_2)~.\label{alpha-in-conf-frame}
\eea
Our next task is to express the $\beta_j$ in terms of geometrical data of the original triangle with circular edges. Let us start with $\beta_3$. 

Since the straight line connecting $y_1$ and $y_2$ extends up to $y_3=\infty$ it is the image of the circumcircle of the original triangle with corners $x_1,x_2,x_3$. Angles of crossing lines are preserved. Therefore $\beta_3$ is the angle both at $x_1$ and $x_2$ between the circumcircle and
the circular edge connecting these two corners.

Beyond this characterisation, we still want to have an expression for $\beta_3$ in terms of distances between the corners  $x_j$ and/or the centers of the circular arcs $z_j$.\footnote{As in the main text using the notation $D_{ij}=\vert x_i-x_j\vert $, but with  footnote \ref{footnote-10} $R_3=\vert x_1-z_3\vert=\vert x_2-z_3\vert$ etc.}
The unit tangential vector at $x_1$ to the circular edge pointing along the circle in the direction of $x_2$ is
\beq
t_{\mbox{\scriptsize edge}}~=~\frac{x_2-z_3-(x_1-z_3)\cos\delta}{R_3\vert\sin\delta\vert}
\eeq
with $\cos\delta=1-\frac{D_{12}^2}{2R_3^2}$.

The unit tangent vector at $x_1$ to the circumcircle pointing along this circle in the direction of $x_2$ is
\beq
t_{\mbox{\scriptsize cc}}~=~\frac{(x_2-x_1)D_{13}^2-(x_3-x_1)D_{12}^2}{D_{12}D_{23}D_{13}}~.
\eeq 
Then we get for $\cos\beta_3=t_{\mbox{\scriptsize edge}}t_{\mbox{\scriptsize cc}}$ after some algebra
\beq
\cos\,\beta_3~=~\frac{R_3}{D_{13}D_{23}\sqrt{4R_3^2-D_{12}^2}}~\left (D_{13}^2+D_{23}^2-\frac{D_{12}^2(R_3^2+(x_3-z_3)^2)}{2R_3^2}\right )~.
\eeq
The other two angles $\beta_1$ and $\beta_2$ are obtained by corresponding cyclic permutations of the indices. 

There is a closure condition for the six conformal invariant angles $\alpha_j,~\beta_j,~j=1,2,3$  
\beq
f(\alpha_1,\dots,\beta_3)~=~0~.\label{closure}
\eeq
It reduces the number of independent conformal invariants to five, according to  the counting based on the  representation in  the conformal frame above.
We did not find a short symmetric expression for the function $f$ in \eqref{closure}. However its solution, expressing e.g. $\alpha_3$ in terms
of the five other angles, can be found by first expressing in the above conformal frame $(\vartheta_1,\varphi_1)$ by 
$(\alpha_1,\beta_2,\beta_3)$ as well as $(\vartheta_2,\varphi_2)$ by $(\alpha_2,\beta_1,\beta_3)$ and putting this into \eqref{alpha-in-conf-frame}.

Finally we express the cusp angles in terms of distances
\bea
\cos\, \alpha_3&=&\frac{2R_1R_2}{D_{13}D_{23}\sqrt{(4R_1^2-D_{23}^2)(4R_2^2-D_{13}^2)}}\\[2mm]
&\cdot &\left (\frac{(x_1-z_1)^2D_{23}^2}{2R_1^2}+\frac{(x_2-z_2)^2D_{13}^2}{2R_2^2}-\frac{(z_1-z_2)^2D_{13}^2D_{23}^2}{4R_1^2R_2^2}\right .\nonumber\\[2mm]
&&\left .~~~~~~~~~~~~~~~~~~~~~~~~~~~~~+\frac{D_{13}^2+D_{23}^2-2D_{12}^2}{2}-\frac{D_{13}^2D_{23}^2}{4R_1^2R_2^2}(R_1^2+R_2^2)\right )~,\nonumber
\eea
and corresponding cyclic permutations.


\end{document}